\documentclass[twocolumn,showpacs,preprintnumbers,amsmath,amssymb,prb]{revtex4-1}

\usepackage[colorlinks=true,bookmarks=false,citecolor=blue,urlcolor=blue]{hyperref} 

\usepackage{amsmath,amssymb}
\usepackage[mathscr]{euscript}
\usepackage{graphicx}
\usepackage{verbatim} 
\usepackage{color}

\usepackage[english]{babel}
\usepackage{bm}
\usepackage{natbib}

\def\dfrac{\displaystyle\frac}  

\renewcommand {\Re}{\mathop{\mathrm{Re}}\nolimits}   

\renewcommand {\phi}{\varphi}

\newcommand{\eps}{\varepsilon}

\begin{document}
\title{Cascaded third harmonic generation in hybrid graphene-semiconductor waveguides}

\author{Daria A. Smirnova}
\author{Alexander S. Solntsev}

\affiliation{
Nonlinear Physics Center, Research School of Physics and Engineering, Australian National University, 
Canberra ACT 0200, Australia}

\pacs{78.67.Wj, 42.65.Wi, 73.25.+i, 78.68.+m}

\begin{abstract} 
We study cascaded harmonic generation of hybrid surface plasmons in integrated planar waveguides composed of a graphene layer and a doped-semiconductor slab. We derive a comprehensive model of cascaded third harmonic generation through phase-matched nonlinear interaction of fundamental, second harmonic and third harmonic plasmonic modes supported by the structure. We show that hybrid graphene-semiconductor waveguides can simultaneously phase-match these three interacting harmonics, increasing the total third-harmonic output by a factor of 5 compared to the non-cascaded regime.
\end{abstract}

\maketitle

\section{Introduction} 

Parametric wavelength conversion in general
and harmonic generation in particular are among the
most intensively studied phenomena in optics
and photonics~\cite{Boyd}. It was demonstrated that important advantages can be obtained by cascading several parametric processes~\cite{Saltiel2005, SolntsevAPL2011}. For instance, since a wide range of materials provide stronger quadratic nonlinear response
compared to cubic nonlinear response, the efficiency of the third harmonic generation (THG) can be strongly enhanced by utilising a cascaded process consisting of second harmonic generation (SHG) and sum-frequency generation (SFG), which mixes fundamental wave (FW) and
second harmonic (SH) to generate the third harmonic (TH).
The conversion efficiency in this regime is the highest when the phase-matching
(PM) conditions for both SHG and SFG are satisfied.

Doped graphene is an atomically thin tunable plasmonic material~\cite{RevGrigorenko, Lozovik_Usp_2012}, which can be incorporated into various components in nanoscale and microscale optics. In particular, graphene demonstrates strong optical nonlinearity, especially when utilised near the plasmon resonance~\cite{ PRL_SinglePhoton, Mikhailov2011_SHG, PRL_Tokman_2014}. It is therefore of significant interest to establish a theoretical framework and explore the viability of cascaded nonlinear optical processes in graphene-enhanced wavelength conversion devices.

\begin{figure}[!b]\centering
  \includegraphics[width=1\linewidth]{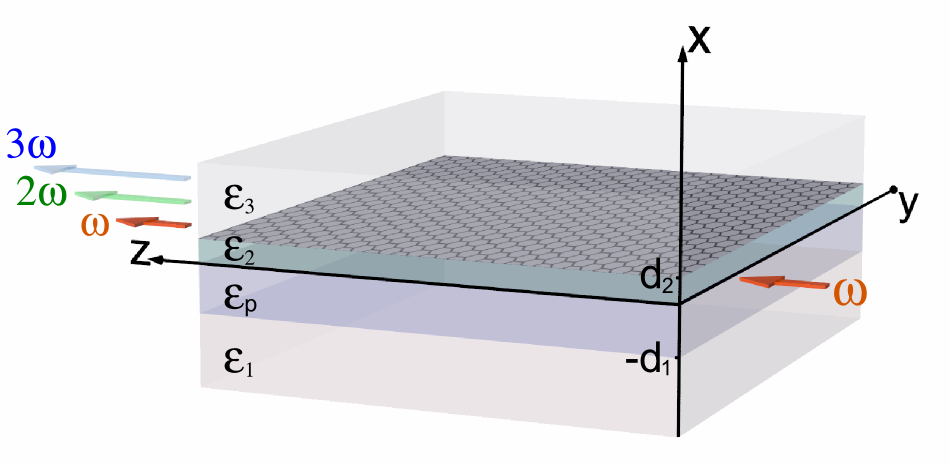} 
  \caption{(Color online) Geometry of a planar hybrid graphene-semiconductor waveguide. A waveguide consists of a monolayer graphene sheet placed at the height $d_2$ above the doped semiconductor with permittivity $\varepsilon_p$ and thickness $d_1$ sandwiched between two dielectric layers with permittivities $\varepsilon_1$ and $\varepsilon_2$. The third dielectric layer with permittivity $\varepsilon_3$ is placed on top of a graphene layer. The optical pump at the frequency $\omega$ (red arrow) is generating the second harmonic at the frequency $2 \omega$ (green arrow) and the third harmonic at the frequency $3 \omega$ (blue arrow).}  
  \label{fig:fig1}
\end{figure}

In this work we develop a theoretical model for cascaded third harmonic generation in graphene-based plasmonic waveguides. We show that engineering modal dispersion allows to simultaneously satisfy SHG and SFG phase-matching for efficient cascaded THG, and offers 5 times increase of TH output compared to direct non-cascaded THG.
We consider a planar waveguiding system 
consisting of a graphene sheet placed over a doped semiconductor sandwiched between two dielectric layers, see Fig.~\ref{fig:fig1}. The dielectric permittivity of a doped semiconductor is described by the Drude model with the plasma frequency comparable to the Fermi energy in graphene.
This geometry allows the coexistence of three plasmonic modes\cite{GrPlasmaLayerWgAPL2013}, which as we show below can be phase-matched, strongly increasing effective nonlinear optical interaction. In our model we assume that the nonlinear interaction of propagating plasmons is achieved due to the nonlinear response in graphene. In this work we focus only on nonlinear processes in the waveguide; the way of linear in-coupling and out-coupling of light to and from such hybrid-graphene waveguides was studied elsewhere~\cite{GrapheneWG_SHG_2014}.

This paper is structured as follows. In Sec. II, we derive the expressions for the nonlinear conductivities at FW, SH and TH in graphene. In Sec. III, we derive the SHG and THG coupled-mode equations and the corresponding nonlinear optical coefficients, including direct and cascaded contributions to the TH. In Sec. IV, we study the phase-matching and present the results of a numerical simulation, showing that hybrid graphene-semiconductor waveguides can simultaneously phase-match these three interacting harmonics, increasing the total third-harmonic output by a factor of 5 compared to non-cascaded regime.

\section{Nonlinear conductivities of graphene}

First, to characterize the nonlinear response of a graphene layer, we methodologically derive expressions for nonlinear conductivities taking into account in-plane plasmon momentum, which results in nonzero quadratic conductivity and enables SHG. 
To describe nonlinear properties of doped graphene, we employ a quasi-classical description based on the solution of the Boltzmann equation~\cite{Mikh_Nonlin, Mikh_Ziegler_Nonlin, Glazov2013, Glazov2011_SHG}. It is applicable at low frequencies, $\hbar \omega \leq \mathcal{E}_{F}$,  where $\mathcal{E}_{F}$ is the Fermi energy. In the collisionless limit, the electron distribution function $f(\mathbf{r},\mathbf{p},t)$ satisfies the kinetic equation written as
\begin{equation}
\frac{\partial f}{\partial t} + \mathbf{v_{\mathbf p}} \frac{\partial
 f}{\partial \mathbf{r}} + e \left( \mathbf{E} + \frac{1}{c} [ \mathbf{v_{\mathbf p}} \times \mathbf{B}] \right) \frac{\partial f}{\partial \mathbf{p}}  = 0\:, 
\end{equation}
where for Dirac's massless quasi-particles $\mathbf{v_{\mathbf p}} = \dfrac{\partial \mathscr{E}}{\partial \mathbf p}  = v_{F}\dfrac{\mathbf p}{|\mathbf p|}$, the energy $\mathscr{E} = v_{F}p$, $v_{F}\approx c/300$ is the Fermi velocity, and $e=-|e|$ is the charge. Following the perturbation approach~\cite{Glazov2011_SHG, Mikhailov2011_SHG}, we solve this equation using iterations.

Here we consider the case of p-plasmons or TM-polarized incident wave with the tangential electric field component of the form $\mathbf{E}_{\omega} = E \mathbf{x}_0 e^{iqx}$ (and $ q v_{F}/ \omega \ll 1$). In this section, we assume that a graphene monolayer is placed in the plane $xy$. We note that there is no component of magnetic field normal in respect to the graphene surface. Because of the graphene nonlinearity, the field $\mathbf{E}_{\omega}$ excites higher harmonics. We look for the distribution function in the form
\begin{equation}
\begin{split}
f = f_0(\varepsilon) +  f^{(2)}_{0} + (f^{(1)}_{\omega} + f^{(3)}_{\omega} ) e^{-i( \omega t - qx )} + \text{c.c.}
\\ + f^{(2)}_{2 \omega}e^{-2i( \omega t - qx )}+ \text{c.c.} + f^{(3)}_{3 \omega}e^{-3i( \omega t - qx )}  + \text{c.c.} +  \ldots \:,
\end{split}
\end{equation}
where $f_0(\mathscr{E})$ is the Fermi-Dirac distribution (step-like at low temperatures, $ k_B T \ll \mathcal{E}_{F}$), $f^{(1)}_{\omega}$, $f^{(2)}_{2 \omega}$, $f^{(3)}_{3 \omega}$ are the first, the second, and the third order oscillating corrections (the superscripts are omitted below), respectively, whereas the tangential electric field is as follows
\begin{equation}
\begin{split}
\mathbf{E}_{\tau} = \mathbf{E}_{\omega} e^{-i( \omega t - qx )} + \text{c.c.}  + \mathbf{E}_{2 \omega} e^{-2i( \omega t - qx )} + \text{c.c.} \\+ \mathbf{E}_{3 \omega} e^{-3i( \omega t - qx )}  + \text{c.c.} + \ldots \:.
\end{split}
\end{equation}

The electric current density at $\omega$ is given by the first order correction to the Fermi-Dirac distribution function
$${\bf j}_{\omega} = 4 e \sum\limits_{{\bf p}} \mathbf{v_{\mathbf p}}  f_{\omega} ({\mathbf p})  = 
 4 e \int \int \dfrac{dp_x dp_y}{(2 \pi \hbar)^2}  v_{p_x} f_{\omega} \mathbf{x}_0  \:,$$ 
where
\begin{equation}
 f_{\omega} = \dfrac{e \left(\dfrac{\partial f_0}{\partial {\mathbf p}} {\mathbf E_{\omega}}\right)} {i(\omega - v_{p_x} q ) }\approx 
 \dfrac{1}{i\omega} \left  ( 1 + \dfrac{v_{p_x} q}{\omega} \right) e E \dfrac{p_x} {p} \dfrac{\partial f_0}{\partial { p}} \:.
\end{equation}
Therefore, the first order current amplitude can be written as follows
\begin{equation}
\begin{aligned}
 j_{\omega_x}
& = 
\int \int 4 e v_{p_x} \left(  \dfrac{eE} {i \omega} \dfrac{p_x} {p}  \dfrac{\partial f_0} {\partial p}  \right)  \dfrac{dp_x dp_y}{(2 \pi \hbar)^2}\\ & =
\int_{0}^{\infty} \dfrac{ p dp}{(2 \pi \hbar)^2} \int_{0}^{2 \pi} d \Theta \left(  \dfrac{4 e^2 E} {i \omega} \dfrac{\partial f_0} {\partial p}  \right) v_F \cos^2{\Theta}\\ 
& = \displaystyle{\frac{ie^2}{\pi \hbar^2}\frac{\mathcal{E}_{F}}{\omega}} E\:.
\end{aligned}
\end{equation}
This is a known result~\cite{Mikh_Nonlin}. 
In the local semiclassical limit, the linear conductivity of graphene can
be reduced to the Drude model, which only takes into account intraband processes~\cite{Mikh_Nonlin}
\begin{equation}
\sigma_{\text{intra}} (\omega)  =  \displaystyle{\frac{ie^2}{\pi \hbar^2}\frac{\mathcal{E}_{F}}{\omega }},
\end{equation}
It can also be corrected to include relaxation
\begin{equation}
\sigma(\omega)  =  \displaystyle{\frac{ie^2}{\pi \hbar^2}\frac{\mathcal{E}_{F}}{\left(\omega + i\tau_{\text{intra}}^{-1}\right)}},
\end{equation}
where $\displaystyle{\tau_{\text{intra}}^{-1}}$ is the relaxation rate. 

The second order correction is defined as follows
\begin{widetext}
\begin{equation} \label{eq:f_2}
\begin{aligned}
& f_{2 \omega} = 
\dfrac{e E \dfrac{\partial f_{\omega}}{\partial { p_x}}  } {2i(\omega - v_{p_x} q)} \approx 
\dfrac{1} {2i\omega}   \left( 1 + \dfrac{v_{p_x} q}{\omega}  \right) e E \dfrac{\partial} {\partial p_x} 
\left(  \dfrac{1}{i\omega} \left  ( 1 + \dfrac{v_{p_x} q}{\omega} \right) e E \dfrac{p_x} {p} \dfrac{\partial f_0}{\partial { p}} \right) \approx \\
  &- \dfrac{1}{2 \omega^2}  e^2 E^2 \dfrac{\partial} {\partial p_x} \left( \dfrac{p_x} {p} \dfrac{\partial f_0}{\partial p} \right) -
\dfrac{1}{2 \omega^2}  e^2 E^2  \left( \dfrac{v_{p_x} q} {\omega}  \dfrac{\partial} {\partial p_x}  \left( \dfrac{p_x} {p} \dfrac{\partial f_0}{\partial p} \right)  +       \dfrac{\partial} {\partial p_x}  \left[\dfrac{v_{p_x} q} {\omega}  \dfrac{p_x} {p} \dfrac{\partial f_0}{\partial p}   \right] \right)
 \:.
\end{aligned}
\end{equation}
\end{widetext}
It gives rise to the electric current density at $2\omega$ 
${\bf j}_{2 \omega} = 4 e \sum\limits_{{\bf p}} \mathbf{v_{\mathbf p}}  f_{2 \omega} ({\mathbf p}) \:.$
Integrating the first term in Eq.~(\ref{eq:f_2}) returns zero. Integrating the second term leads to the following expression
\begin{widetext}
\begin{equation}
\begin{aligned}
& {j}_{2 \omega_x }  =  - 4 e \left( \dfrac{e^2 E^2} {2 \omega^2} \right) \dfrac{q} {\omega}
\int_{0}^{\infty} \dfrac{ p dp}{(2 \pi \hbar)^2} \int_{0}^{2 \pi} d \Theta v_F \cos \Theta \\
& \times \Biggl [ v_F \cos \Theta \left( \cos \Theta \dfrac{\partial} {\partial p}  - \dfrac{\sin \Theta}{p} \dfrac{\partial} {\partial \Theta} \right) \left(\cos \Theta \dfrac{\partial f_0} {\partial p} \right) + 
\left( \cos \Theta \dfrac{\partial} {\partial p}  - \dfrac{\sin \Theta}{p} \dfrac{\partial} {\partial \Theta} \right) \left(v_F \cos^2 \Theta \dfrac{\partial f_0} {\partial p} \right) \Biggr]\:,
\end{aligned}
\end{equation}
\end{widetext}
which can be reduced to
\begin{equation}
{j}_{2 \omega_x }=  - \dfrac{3} {8} \dfrac{e^3 v_F^2} {\pi \hbar^2 \omega^3} q E^2 \:.
\end{equation}
Therefore we obtain the nonlinear source
\begin{equation}
{\bf j}_{2 \omega} =  - \mathbf{x}_0 \dfrac{3} {8} \dfrac{e^3 v_{F}^2} {\pi \hbar^2 \omega^3} q E^2 \exp (-i2 \omega t + i2qx) \:,
\end{equation}
which is in agreement with the earlier results~\cite{Mikhailov2011_SHG,SHG_Wrapped}.
Thus, considering frequency conversion with plasmons, the quadratic conductivity of graphene $\sigma_{2} (\omega)$ can be defined as follows
\begin{equation}
\sigma_{2} (\omega) = -  \dfrac{3} {8} \dfrac{e^3 v_{F}^2} {\pi \hbar^2 \omega^3} q,
\end{equation}
where $q=k_{\text{sp}} (\omega)$ is the plasmon wavenumber supported by a waveguide at $\omega$. 

Next, the third order correction is defined as
\begin{equation} \label{eq:f_3}
\begin{aligned}
f_{3 \omega} & = 
\dfrac{1  } {3i(\omega - v_{p_x} q)} 
\left( e E_{\omega}\dfrac{\partial f_{2\omega}}{\partial { p_x}}  +   e E_{2\omega}\dfrac {\partial f_{\omega}}{\partial { p_x}} \right) \\
&  \equiv f^{\text{(I)}}_{3 \omega} + f^{\text{(II)}}_{3 \omega} 
\:.
\end{aligned}
\end{equation} 
To obtain the contribution to the third harmonic current ${\bf j}_{3 \omega} = 4 e \sum\limits_{{\bf p}} \mathbf{v_{\mathbf p}}  f_{3 \omega} ({\mathbf p})$ from the first term 
\begin{equation}
f^{\text{(I)}}_{3 \omega} = 
\dfrac{1  } {3i(\omega - v_{p_x} q)}  e E_{\omega}\dfrac{\partial f_{2\omega}}{\partial { p_x}} , 
\end{equation}
we expand this term as follows 
\begin{widetext}
\begin{equation}
\begin{aligned}
& f^{\text{(I)}}_{3 \omega} 
 =  \dfrac{1} {3i\omega}   \left[ 1 + \dfrac{v_{p_x} q}{\omega}  + \left(\dfrac{v_{p_x} q}{\omega}\right)^2 \right]\\
& \times e E_{\omega} \dfrac{\partial} {\partial p_x} 
\left[  \dfrac{1}{2i\omega} \left  [ 1 + \dfrac{v_{p_x} q}{\omega} + \left(\dfrac{v_{p_x} q}{\omega}\right)^2 \right] e E_{\omega}  \dfrac{\partial} {\partial p_x} 
\left(  \dfrac{1}{i\omega} \left  [ 1 + \dfrac{v_{p_x} q}{\omega} + \left(\dfrac{v_{p_x} q}{\omega}\right)^2 \right] e E_{\omega} \dfrac{p_x} {p} \dfrac{\partial f_0}{\partial { p}} \right)    \right]\:. 
\end{aligned}
\end{equation}
\end{widetext}
Here integrating the expressions proportional to $q^1$ returns zero.
After integrating the terms proportional to $q^0$ and $q^2$, the current amplitude can be written as follows
\begin{equation}
j^{\text{(I)}}_{3 \omega_x} = \sigma_3(\omega )  E_{\omega}^3 = \left[ \sigma_3(\omega, q=0)  + \Delta \sigma_3(\omega, q) \right] E_{\omega}^3 \:, 
\end{equation}
where the third order conductivity, in the homogeneous case~\cite{ Mikh_Ziegler_Nonlin, Mikh_Nonlin, Glazov2013, Peres2014} given by   
\begin{equation}
\sigma_3 (\omega, q=0) = i \displaystyle{\frac{1}{8} \frac{e^2}{\pi \hbar^2} \left(\frac{ e v_{F} } { \mathcal{E}_{F} \omega}\right)^2 \frac{ \mathcal{E}_{F} } { \omega}} =
i \displaystyle{\frac{1}{8} \frac{e^4}{\pi \hbar^2} \frac{  v_{F} ^2} { \mathcal{E}_{F} \omega^3}} \:, 
\end{equation}
is corrected by an additional term  
\begin{equation}
\begin{aligned}
\Delta \sigma_3 (\omega, q) & =
i \displaystyle{\frac{5}{16} \frac{e^4}{\pi \hbar^2} \frac{  v_{F} ^2} { \mathcal{E}_{F} \omega^3}} \left( \dfrac{q v_F}{\omega} \right)^2 \\
& = \dfrac{5}{2}\sigma_3 (\omega, q=0) \left( \dfrac{q v_F}{\omega} \right)^2  \:. 
\end{aligned}
\end{equation}

Further, by analogy 
with the second harmonic contribution $f_{2\omega}$, we derive the third harmonic current corresponding to the second term $f^{\text{(II)}}_{3 \omega}$ given by 
\begin{equation}
\begin{aligned}
f^{\text{(II)}}_{3 \omega} & = \dfrac{1  } {3i(\omega - v_{p_x} q)} 
 e E_{2\omega}\dfrac{\partial f_{\omega}}{\partial { p_x}} = \dfrac{1} {3i\omega}   \left( 1 + \dfrac{v_{p_x} q}{\omega}  \right) \\
 & e E_{2\omega} \dfrac{\partial} {\partial p_x} 
\left(  \dfrac{1}{i\omega} \left  ( 1 + \dfrac{v_{p_x} q}{\omega} \right) e E_{\omega} \dfrac{p_x} {p} \dfrac{\partial f_0}{\partial { p}} \right)\:.
\end{aligned}
\end{equation}
and get the x-projection of the nonlinear current as follows
\begin{equation}
j^{\text{(II)}}_{3 \omega_x}=\dfrac{2}{3} \sigma_{2} (\omega) E_{2\omega} E_{\omega}\:,   
\end{equation}
which entirely corresponds to the cascaded process in THG.

\section{Coupled mode equations} 

Neglecting losses, the dispersion relation and p-polarized eigenmodes of a hybrid graphene-semiconductor waveguide were analysed in Ref.~\cite{GrPlasmaLayerWgAPL2013}. 
Similar to the approach outlined in Ref.~\cite{GrapheneWG_SHG_2014}, to describe the plasmon conversion due to the graphene nonlinearity, we 
derive equations for the modal envelopes. The frequency conversion can be modeled by coupled equations for the amplitudes of FW, SH and TH plasmonic modes, $\mathcal{A}_1$, $\mathcal{A}_2$ and $\mathcal{A}_3$, respectively. Assuming that the conversion efficiency is small, we utilize the following magnetic field ansatz 
\begin{equation}
\begin{aligned}
& H_y(x,z,t) =  e^{-i\omega t} \mathcal{A}_1(\mu z) h_1(\omega,x)e^{ik^1_{\text{sp}}(\omega)z} \\
& + 
e^{-i 2\omega t} \mathcal{A}_2(\mu z) h_2(2\omega,x) e^{ik^2_{\text{sp}}(2\omega)z}
 \\
&  + e^{-i 3\omega t} \mathcal{A}_3(\mu z) h_3(3\omega,x)e^{ik^3_{\text{sp}}(3\omega)z}\:,
\end{aligned} 
\end{equation}
where $h_{1,2,3}(x)$ are the transverse profiles of the modes.   
Disregarding the influence of the generated modes on the pump, the coupled-mode equations can be written as follows
\begin{subequations} \label{eq:CMT}
\begin{align} 
& 
\dfrac{d \mathcal{A}_1} {d z} + \gamma_1 \mathcal {A}_1  = 0 \:,\\
& 
\dfrac{d \mathcal{A}_2} {d z} + \gamma_2 \mathcal {A}_2 = g_2  \mathcal{A}_1^2 e^{-i [  k^2_{\text{sp}}(2\omega) - 2 k^1_{\text{sp}}(\omega) ]z}\:, \\ 
& 
\dfrac{d \mathcal{A}_3} {d z} + \gamma_3 \mathcal {A}_3  = \bar{g}_2  \mathcal{A}_1 \mathcal{A}_2 e^{-i [  k^3_{\text{sp}}(3\omega)  -  k^2_{\text{sp}}(2\omega) -  k^1_{\text{sp}}(\omega) ]z} \\ \nonumber
& \quad \quad \quad \quad \quad \quad  + {g}_3  \mathcal{A}_1^3  e^{-i [  k^3_{\text{sp}}(3\omega)  -  3 k^1_{\text{sp}}(\omega) ]z} \:, 
\end{align}
\end{subequations}
where the term proportional to $\bar{g}_2 $ is responsible for the so-called  {\em cascaded} contribution to the THG~\cite{SolntsevAPL2011,Saltiel2005}.
Assuming that the FW plasmon is launched at $z=0$, we study the plasmon frequency conversion modeled by Eqs.(\ref{eq:CMT}). The conversion from the TH to the SH can be neglected, because, as will be shown below, the TH efficiency in this kind of graphene-semiconductor waveguides remains much smaller than the SH efficiency, $\mathcal{A}_3(z) \ll \mathcal{A}_2(z)$. 

Loss coefficients $\gamma_{1,2,3}$ in Eqs.(\ref{eq:CMT}) can be obtained   perturbatively from the dispersion equation, introducing dissipative corrections, namely the linear frequency-dependent surface conductivity of graphene 
\begin{equation}\sigma (\omega ) \equiv \sigma^{(R)} (\omega )  + i \sigma^{(I)}(\omega ),
\end{equation}
and the dielectric permittivity of the semiconductor 
\begin{equation}
\eps_ p(\omega) = 1 - \dfrac{\omega_p^2}{\omega(\omega+i\nu)} = 1 - \dfrac{\omega_p^2}{\omega^2} + \dfrac{i \nu \omega_p^2}{\omega^3},  
\end{equation}
where $\nu$ is a damping coefficient.
The coefficients $g_2$, $\bar{g}_2$, $g_3$ in nonlinear sources can be derived from the equations for the energy flows in the corresponding modes as follows
\begin{subequations}
\begin{align}
& \dfrac{d P^{2}_z}{d z} = - \dfrac{1} {2} \Re \int_{-\infty}^{+\infty} { {\bf{j}}^{*}_{2\omega} {\bf E}_{2\omega} dx} \;,\\
& \dfrac{d P^{3}_z}{d z} = - \dfrac{1} {2} \Re \int_{-\infty}^{+\infty} { {\bf{j}}^{*}_{3\omega} {\bf E}_{3\omega} dx}\;,
\end{align}
\end{subequations}
where $ P^{2,3}_z = G_{2,3} |\mathcal{A}_{2,3}|^2$, $G_{2,3}$ are defined as 
\begin{equation}
\begin{aligned}
G_{2,3} & = \Re \dfrac{c}{8 \pi} \int_{-\infty}^{+\infty} \left[ {\bf e}_{2,3} {\bf h}_{2,3}^{*} \right] dx 
\\ & =  \Re \dfrac{c}{8 \pi} \int_{-\infty}^{+\infty} \left[ { e^{2,3}_x(x)} {h^{2,3 *}_y(x)}  \right] dx  \:. 
\end{aligned}
\end{equation}
Here $e^{2,3}_x(x)$ and  ${h^{2,3 }_y(x)}$ are the transverse profiles of the SH and TH modes normalized in such a way that at the graphene layer $e_z(d_2) = 1$. The field components at the p-polarized surface wave are connected as 
\begin{equation}
e_z(x) = \dfrac{1}{-ik_0\eps(x)} \dfrac{d h_y}{dx}= \dfrac{1}{-ik_{\text{sp}}} \dfrac{d e_x}{dx}. 
\end{equation}
Finally, we can write the nonlinear coefficients as follows 
\begin{equation}
g_2 = - \dfrac{\sigma_2}{4 G_2}, \quad \bar g_2 = - \dfrac{\dfrac{2}{3}\sigma_2}{4 G_3}, \quad g_3 = - \dfrac{\sigma_3}{4 G_3}\:.
\end{equation}
These coefficients define the strength of the nonlinear interactions between FW, SH and TH plasmons. Phase matching, another important parameter that defines the conversion between these harmonics, is discussed in the following section.
%
%

\begin{figure}[t]
\centering\includegraphics[width=0.93\linewidth]{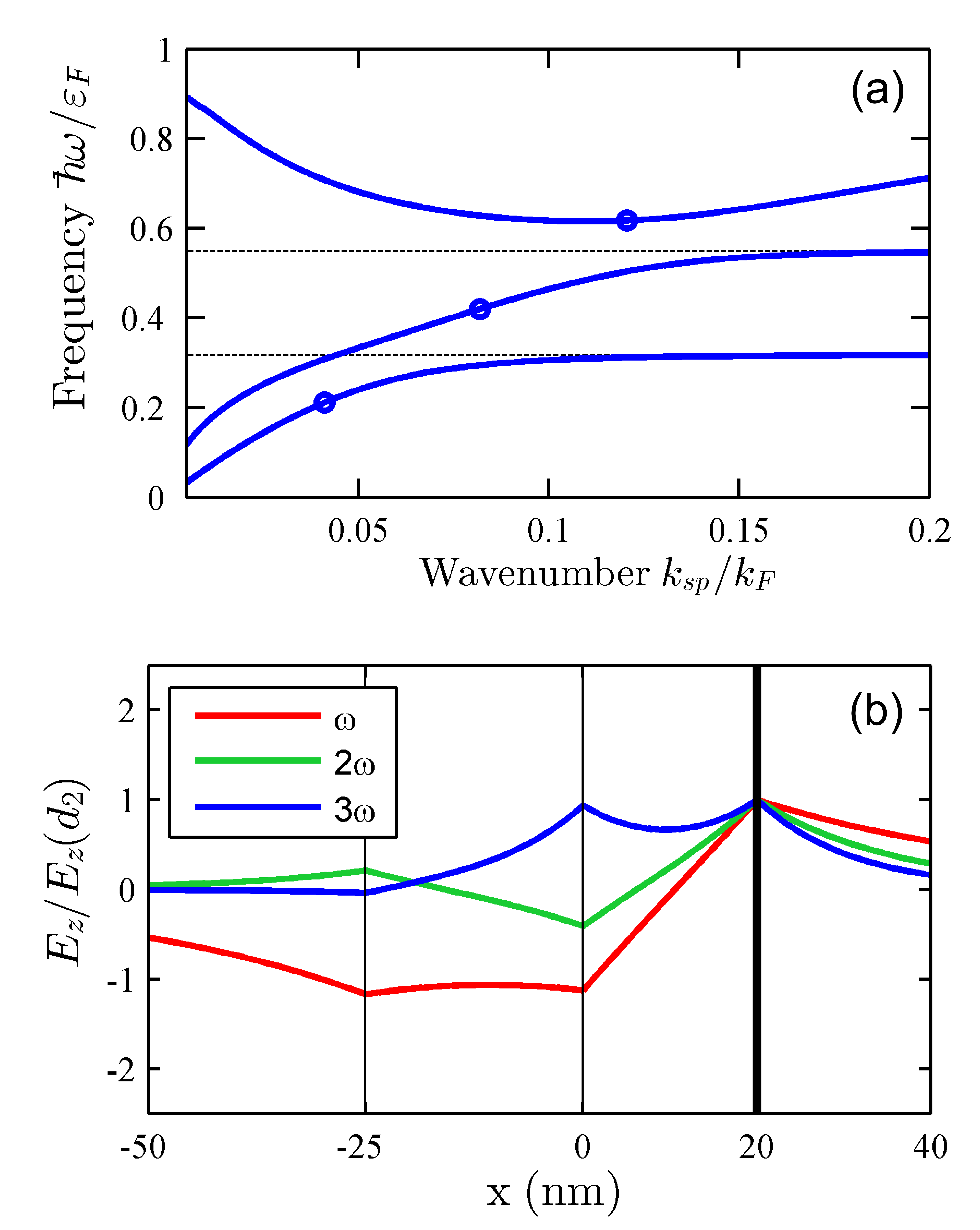} 
\caption{(Color online) (a) Dispersion of three supported plasmon modes in a hybrid graphene-semiconductor waveguide (solid blue lines), corresponding asymptotes (dashed black lines), corresponding points of phase-matching for FW, SH and TH (blue circles);  
(b) FW, SH and TH mode profiles shown for the continuous in-plane electric field component. Bold black line corresponds to the graphene layer, fine black lines correspond to the interfaces between semiconductor/dielectric layers.
}
\label{fig:fig2}
\end{figure}

\section{Phase-matching and wavelength conversion}
Next we show that three guided modes corresponding to FW, SH and TH can be phase-matched, so that at some $\omega$, dependent on the geometrical parameters (separations between the layers, dielectric permittivity distribution and doping levels), they have almost equal effective indexes $\beta^1({\omega}) \approx \beta^2(2\omega) \approx \beta^3(3\omega) $ (or, equivalently, $k^1_{\text{sp}}(\omega) = k^2_{\text{sp}}(2\omega)/2 = k^3_{\text{sp}}(3\omega)/3 $). To calculate the dispersion and the mode profiles we use the transfer matrix method~\cite{Born}.

\begin{figure}[!t]
\centering\includegraphics[width=0.93\linewidth]{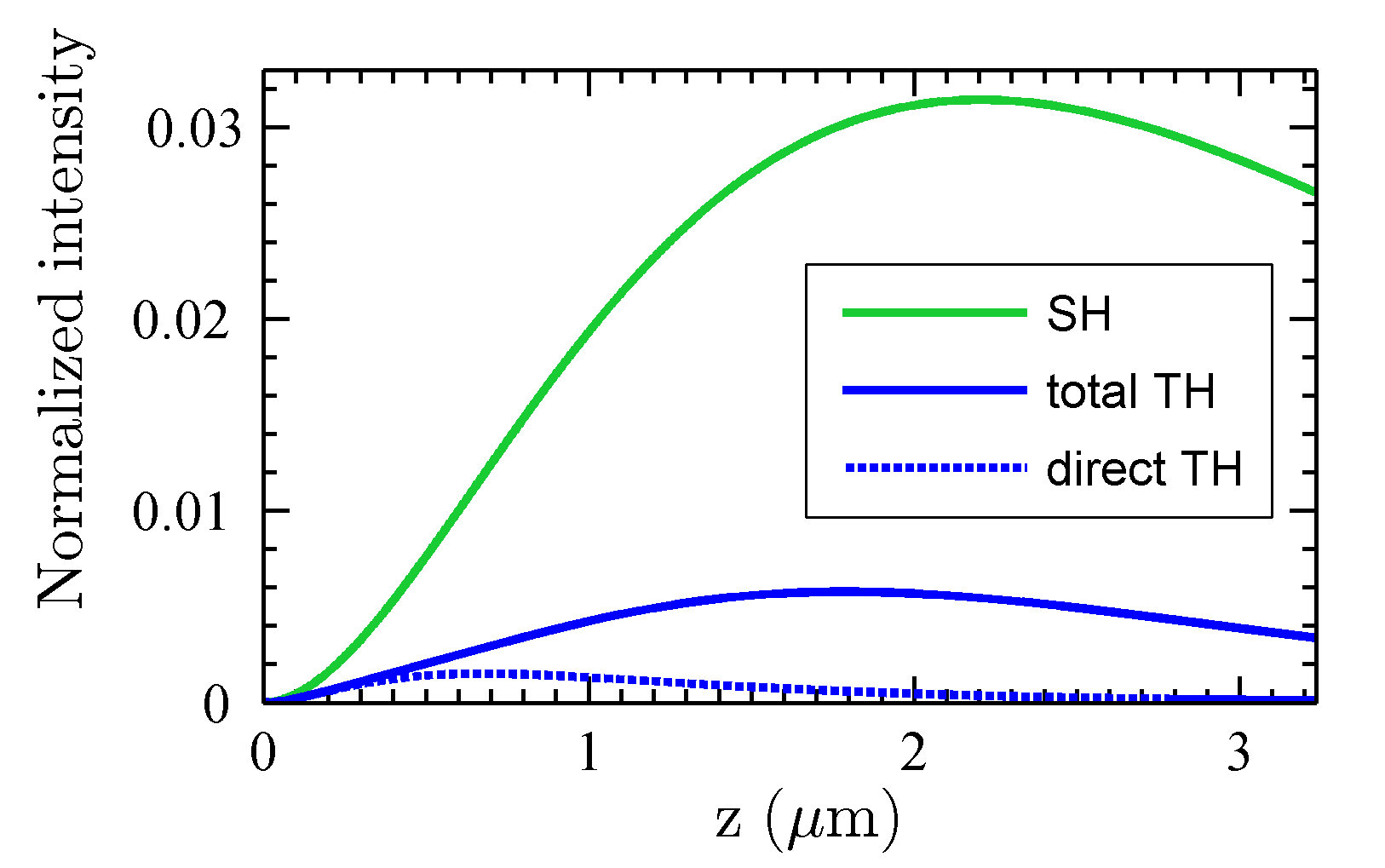} 
\caption{(Color online) Spatial evolution along the propagation direction of the normalized SH intensity $|\mathcal{A}_2(z)|^2/|\mathcal{A}_1(0)|^2$ (solid green line), total TH intensity including direct THG and cascaded THG $|\mathcal{A}_3(z)|^2/|\mathcal{A}_1(0)|^2$ (solid blue line), and the direct TH intensity $|\mathcal{A}_3(z)|^2/|\mathcal{A}_1(0)|^2$ in the absence of the cascading effect assuming $\bar g_2 = 0$ (dashed blue line). 
}
\label{fig:fig3}
\end{figure}

To demonstrate this behaviour we choose the following set of realistic waveguide parameters. The permittivities of the dielectric layers shown in Fig.~\ref{fig:fig1} are chosen as $\varepsilon_1=8$, $\varepsilon_2=2$, and $\varepsilon_3=1$. For simplicity we do not take into account the dispersion in the dielectric layers. Adding the dielectric dispersion will require adjusting the thicknesses of the dielectric layers to compensate for it. We choose the thickness of a semiconductor layer as $d_1 = 25$ nm, and the thickness of the dielectric layer between the semiconductor and graphene as $d_2=20$ nm. Then for the Fermi energy $\mathcal{E}_{F} = 0.5$ eV, plasma  frequency $\omega_p = 0.95 \mathcal{E}_{F}$, relaxation parameter $\tau_{\text{intra}} = 0.3$ ps and the damping coefficient $\nu = 0.001 \omega_p$, we obtain simultaneous phase-matching for FW, SH and TH, as demonstrated in Fig.~\ref{fig:fig2}(a), for the fundamental wavelength in free space of $11.7$ $\mu$m.
In Fig.~\ref{fig:fig2}(b) we show the corresponding overlap between the FW, SH and the TH modes. Strong graphene nolinearity allows efficient nonlinear interaction even between the modes with moderate field confinement, which can be further improved by optimizing the waveguide parameters. Remarkably, for different dielectric and semiconductor parameters, the TH mode can be phase-matched in the backward direction, although in this paper we focus on the case when all three modes are phase-matched in the forward direction. 

In Fig.~\ref{fig:fig3} we demonstrate the evolution of the interacting FW, SH and TH plasmons. We choose a realistic pump intensity of 1 MW/cm$^2$ on a graphene layer~\cite{damage_Krauss_APL_2009, damage_Currie_APL_2011, damage_Roberts_APL_2011, damage_Kiisk_ASS_2013}. The resulting SHG and THG conversion efficiencies and optimal conversion lengths are comparable to the previous works~\cite{PRL_Tokman_2014,GrapheneWG_SHG_2014,Mikhailov_QTHG_2014}. 
Particularly, choosing $y$ dimension of $1$ cm and the pump Poynting flux per unit length $1$ W/cm, the outcome energy flows $P^{2,3}_z$ reach $2\times10^{-3}$ and $3\times10^{-5}$ W/cm, respectively, at the distance of about $2$ $\mu$m.   
We note that the direct 
contribution to THG governed by the term proportional to $g_3$ in Eq.(\ref{eq:CMT}c) is an order of magnitude smaller than the cascaded one. To show this explicitly, as a dashed line in Fig.~\ref{fig:fig3}, we plot the TH intensity evolution calculated by formally setting $\bar{g}_2=0$. The difference is remarkable, because the plasmon wavevector can notably exceed that in free space, leading to the efficent SHG process stimulating a cascaded conversion to the third harmonic plasmon.

\section{Conclusions}

To conclude, in this work we have established the theoretical framework for the direct and the cascaded THG in graphene-semiconductor waveguides, including a detailed derivation of all coefficients in the coupled-mode equations based on the waveguide parameters. We have shown that these waveguides can support simultaneous phase-matching of FW, SH and TH plasmon modes, and that the cascaded THG can be several times stronger than the direct THG.

\section*{ACKNOWLEDGMENTS}

The authors are grateful to Y.S. Kivshar, M.M. Glazov and L.E. Golub for their comments and suggestions. This work was supported by the Australian National University.


\end{document}